\documentclass[useAMS,usenatbib,usegraphicx]{mn2e}
\usepackage{epsfig}
\usepackage{amsmath} 
\usepackage{rotating}           
\usepackage{color}     
\usepackage{graphicx}
\usepackage{times}
\usepackage{upgreek} 
\usepackage{amssymb} 

\def\kms{km ${\rm s}^{-1}$}
\def\Lo{L$_\odot$}

\def\scm  {$\hbox{{\rm cm}}^{-2}$}    
\def \AL {$\alpha $}     
\def \HI {H{\sc \,i}}
\def \WpHz {W Hz$^{-1}$}
\def\MOLH {\hbox{${\rm H}_2$}}  

\def\lapp{\ifmmode\stackrel{<}{_{\sim}}\else$\stackrel{<}{_{\sim}}$\fi}
\def\gapp{\ifmmode\stackrel{>}{_{\sim}}\else$\stackrel{>}{_{\sim}}$\fi}

\title[Absorption in Redshifted Radio Sources]{Atomic and Molecular Absorption in Redshifted Radio Sources}  
\author[S. J. Curran,  et al.]{S. J. Curran$^{1}$\thanks{Stephen.Curran@vuw.ac.nz},  M. T. Whiting$^{2}$,  J. R. Allison$^{2}$, A. Tanna$^{3}$, E. M. Sadler$^{4}$  and R. Athreya$^{5}$\\
$^{1}$School of Chemical and Physical Sciences, Victoria University of Wellington, PO Box 600, Wellington 6140, New Zealand\\
$^{2}$CSIRO Astronomy and Space Science, PO Box 76, Epping NSW 1710, Australia\\
$^{3}$School of Physics, University of New South Wales, Sydney NSW 2052, Australia\\
$^{4}$Sydney Institute for Astronomy, School of Physics, The University of Sydney, NSW 2006, Australia\\
 $^{5}$Indian Institute of Science Education and Research, 900, NCL Innovation Park, Dr Homi Bhabha Road Pune, Maharashtra 411008, India}

\begin{document}

\date{Accepted ---. Received ---; in original form ---}

\pagerange{\pageref{firstpage}--\pageref{lastpage}} \pubyear{2017}

\maketitle

\label{firstpage}

\begin{abstract}
  We report on a survey for associated \HI\ 21-cm and OH 18-cm absorption with the Giant Metrewave Radio Telescope at
  redshifts $z\approx0.2-0.4$. Although the low redshift selection ensures that our targets are below the critical
  ultra-violet luminosity ($L_{\rm UV}\sim10^{23}$ \WpHz), which is hypothesised to ionise all of the neutral gas in the
  host galaxy, we do not obtain any detections in the six sources searched. Analysing these in context of the previous
  surveys, in addition to the anti-correlation with the ultra-violet luminosity (ionising photon rate), we find a
  correlation between the strength of the absorption and the blue -- near-infrared colour, as well as the radio-band
  turnover frequency. We believe that these are due to the photo-ionisation of the neutral gas, an obscured sight-line being
  more conducive to the presence of cold gas and the compact radio emission being better intercepted by the absorbing
  gas, maximising the flux coverage, respectively.  Regarding the photo-ionisation, the compilation of the previous surveys
  increases the significance of the critical ionising photon rate, above which all of the gas in the host galaxy is
  hypothesised to be ionised ($Q_\text{\HI}\approx3\times10^{56}$~sec$^{-1}$), to $>5\sigma$. This reaffirms that this is
  an ubiquitous effect, which has profound implications for the detection of neutral gas in these objects with the
  Square Kilometre Array.
\end{abstract} 


\section{Introduction}
\label{intro}

Redshifted \HI\ 21-cm absorption can provide an excellent probe of the contents and nature of the early Universe,
through surveys which are not subject to the same flux and redshift constraints suffered by optical studies. For 
instance, measurement of the contribution of the neutral gas content to the mass density of the Universe
at redshifts ($z\lapp1.7$), where the Lyman-\AL\ transition is not accessible to
ground-based optical telescopes. This corresponds to the past 10 Gyr of cosmic history, which are of particular interest
since this is when star formation was at its most vigorous and the stellar mass density in the Universe increased by
more than a factor of four \citep{hb06}.
There is, however,  mounting evidence that the star formation history is not best traced by {\em intervening} galaxies, detected in the Lyman-\AL\ 
absorption of a background continuum source:
\begin{enumerate}
\item While the star formation history exhibits a strong evolution  (e.g. \citealt{hb06,lbz+14}), the neutral gas content in 
damped Lyman-$\alpha$ absorption systems (DLAs)\footnote{Intervening systems where the neutral hydrogen column density exceeds $N_\text{\HI}=2\times10^{20}$~\scm.}, remains approximately constant with look-back time (e.g. \citealt{rtn05,bra12}).
\item Very few DLAs have been detected in Lyman-\AL\ or H\AL\ emission \citep{mff04,fll+10,fln+11,nlp+12,pbk+12}, both
  tracers of  star formation.
\item The heavy element content of DLAs  does not appear to be caused by star formation within the absorbers themselves, but possibly deposited via winds from nearby galaxies \citep{fm14}.
\item The relative paucity of detections of \HI\ 21-cm absorption in high redshift DLAs appears to be dominated by
extrinsic line-of-sight effects, rather than by any intrinsic evolution  (\citealt{cur12,cdda16}). 
\end{enumerate}
Therefore, in order to fully investigate any relation between the evolution of cold neutral gas and the star formation history,
we should also quantify the population of sources in which the absorption is {\em associated} with the continuum 
source itself.

The detection of \HI\ 21-cm is of further interest as the comparison of its redshift with that of other transitions has
the potential to measure past values of the fundamental constants of nature (\citealt{twm+05,tmw+06}), which
may exhibit a spatial \citep{bfk+10,wkm+10}, as well as a temporal \citep{mwf03} variation.  Radio data can yield at
least an order of magnitude in precision in the measurement of the constants over the optical data (see
\citealt{cdk04}).  Furthermore, the OH radical can not only provide measurements of various combinations of constants
through comparison with optical, \HI\ or rotational (millimetre) transitions, but inter-comparison of the hyperfine
18-cm transitions can remove the systematics introduced by the possible velocity offsets present between species
\citep{dar03}. However, the number of redshifted OH absorption systems remains a paltry five, all of which are at a
redshifts of $z\leq0.89$ (\citealt{wc94,wc95,wc96b,wc98,cdn99,kc02a,kcdn03,kcl+05}).
 
In addition to these, molecular absorption has also been detected in 25 DLAs, through \MOLH\ vibrational transitions
redshifted into the optical band at $z\gapp1.7$ (compiled in \citealt{sgp+10} with the addition of
\citealt{rbql03,fln+11,gnp+12,sgp+12,nsr+15,nkb+16}).  However, extensive millimetre-wave observations have yet to
detect absorption from any rotational molecular transition (e.g. \citealt{cmpw03}), leading us to suspect that the
choice of optically bright objects selects against dusty environments, which are more likely to harbour molecules in
abundance.  This is apparent when one compares the DLAs in which H$_2$ has been detected, which have molecular fractions
${\cal F}\equiv\frac{2N_{\rm H_2}}{2N_{\rm H_2}+N_{\rm HI}}\sim10^{-7}-0.3$ and optical -- near-infrared colours of
$V-K\lapp4$ \citep{cwc+11}, with the five known OH absorbers, where ${\cal F}\approx0.7-1.0$ and $V-K\geq4.80$
\citep{cwm+06}. This is a strong indicator that the reddening is due to dust, which protects the molecular gas from the
ambient ultra-violet radiation.

Therefore by selecting very red objects, with colours of $V-K\gapp5$, we may expect OH column densities which can be
detected with current large radio telescopes. However, there exists an additional constraint when searching for
absorption by gas in the hosts of radio galaxies and quasars -- specifically, that neutral hydrogen has never been
detected where the ultra-violet ($\lambda = 912$~\AA) luminosity of the source exceeds $L_{\rm UV}\sim10^{23}$ \WpHz.
This ``critical'' luminosity applies to all redshifts for various heterogeneous (unbiased) samples, as seen through the
non-detection of \HI\ 21-cm absorption at $L_{\rm UV}\gapp10^{23}$ \WpHz\ by
\citet{cww+08,cwm+10,cwsb12,cwt+12,caw+16,ace+12,gmmo14,akk16,gdb+15}.  This is interpreted as the flux limited optical
spectroscopic surveys, which yield the redshift, selecting the most ultra-violet luminous objects at high redshift
\citep{cww+08}, where the corresponding ionising photon rates of $Q_\text{\HI}\gapp3\times10^{56}$~sec$^{-1}$ are
sufficient to ionise all of the neutral gas in a large spiral galaxy \citep{cw12}.\footnote{\citet{akp+17} have recently
  detected \HI\ 21-cm absorption at $z = 1.223$ in TXS\,1954+513, which they claim has an ultra-violet luminosity of
  $L_{\rm UV}= 4\times10^{23}$ \WpHz. Although unreliable, this luminosity is still consistent with the critical value
  (see Sect. \ref{sec:os}).}

At $z\gapp1$ the vast majority of radio sources for which redshifts are available are believed to have luminosities
above the critical value (see figure 4 of \citealt{msc+15}). Thus, in order to increase the associated absorption
statistics, we can select from the large population of radio sources of known redshift at $z\lapp1$ with
$B\gapp17$.\footnote{The $B$ magnitudes are generally more available than the $V$, as well as being given in the Parkes
  Half-Jansky Flat-spectrum Sample catalogue \citep{dwf+97}.}  This should yield UV luminosities below the critical
$L_{\rm UV}\sim10^{23}$ \WpHz\ in the 1400 MHz band ($z\lapp0.4$, see figure 1 of \citealt{cwt+12}).  This magnitude
selection also has the advantage of giving large blue -- near-infrared colours, where we may expect  $B -K \approx6-10$
for the five known OH absorbers, on the basis of their optical -- near-infrared colours of $V
-K \approx5-9$ \citep{cwm+06}.

From the first part of this survey, we \citep{cwwa11} obtained one detection from four targets, three of which were unaffected by RFI,
following which we were awarded further observing time on the  Giant Metrewave Radio Telescope (GMRT) to complete
the remainder of the requested sample. Here we present our results and discuss their implications: In Sect. \ref{obs} we describe
the sample selection, observations and analysis, in Sect. \ref{sec:res} we present our results, in Sect. \ref{sec:disc} we discuss
these in context of the previous \HI\ 21-cm searches and in Sect. \ref{sec:conc} we present our conclusions.

\section{Observations and analysis}
\label{sec:obs}

\subsection{Sample selection}
\label{ss}

In order to obtain a $z\lapp1$ sample of \HI\ and OH absorbers, we selected sources for which both transitions would be
redshifted into the 1420 MHz receiver band (which spans 1000--1450 MHz).  To ensure sufficient flux against which to
detect the absorption (Sect. \ref{sec:obs}), sources were selected from the Parkes Half-Jansky Flat-spectrum Sample
(PHFS, \citealt{dwf+97,fww00}), giving a total of ten targets with redshifts of $z=0.219-0.405$.  Lastly, the sources
were prioritised by faintness, for which we chose $B\gapp19$ (as quoted in the PHFS, Table \ref{targets}), since this
gave the ten faintest targets for which the estimated flux density at the redshifted  \HI\ 21-cm absorption frequency, $S_{\rm
  est}$, was confirmed to exceed 0.5 Jy.
\begin{table*} 
\centering
\begin{minipage}{175mm}
  \caption{The target list by IAU name (the NED names are given in Sect. \ref{sec:obs}). $z$ is the optical redshift of
    the source from the PHFS (see  Sect. \ref{sec:obs}), $B$ [mag] the blue magnitude quoted in the PHFS, $B-K$
    [mag] the corresponding blue -- near-infrared colour, followed by the values derived from our SED fitting. 
As stated in the main text, the PHFS $B$ magnitudes may the erroneous by $>1$ mag and the uncertainties in the $K$
magnitudes, from 2MASS, range from 0.06 -- 0.12 mags.
 The last five columns give the flux density at the redshifted \HI\ 21-cm absorption frequency, $S_{\rm fit}$, the rest-frame
    radio-band turnover frequency, $\nu_{_{\rm TO}}$, the spectral index, where $S\propto \nu^{\alpha}$, the estimated
    $\lambda = 912$~\AA\ monochromatic luminosity, $L_{\rm UV}$, and the ionising photon rate, $Q_\text{\HI}$, from our fits (see
    Fig.~\ref{SEDs}). All quoted uncertainties are derived from the residuals to the fits.}
\begin{tabular}{@{}l c  c c c  cc  c c c   c c@{}} 
\hline
\smallskip
Source    &  $z$  & \multicolumn{2}{c}{PHFS magnitudes} & \multicolumn{2}{c}{Derived magnitudes} & $S_{\rm fit}$ & $\nu_{_{\rm TO}}$& \AL\ & $L_{\rm UV}$ & $\log_{10}Q_\text{\HI}$ \\
               &         &  $B$ [mag]& $B-K$&   $B$  [mag]& $B-K$  &       [Jy]         &  [MHz]  & &  [\WpHz]               &             [sec$^{-1}$]   \\
\hline
0114+074  & 0.343 & 22.14 & 6.75 &  --- & ---  &$1.98\pm0.11$ & $93\pm1$ & $-0.59\pm0.11$ & --- & --- \\ 
0137+012   & 0.260 & 19.44 & 5.62 &$17.64\pm1.21$ & $3.95\pm0.27$ & $1.42\pm0.17$ & --- & $-0.64\pm0.17$ & $1.5\times10^{22}$ & $55.4\pm0.3$ \\ 
0240--217 & 0.314 &19.05 & 5.57 & $19.17\pm0.73$ & $4.83\pm0.18$  & $1.23\pm0.04$ & $762\pm3$ & $-0.12\pm0.04$ & $2.8\times10^{21}$ & $54.4\pm0.2$ \\ 
0454+066 & 0.405 & 19.79 & 5.37 & $18.69\pm0.20$  & $3.59\pm0.04$  & $0.65\pm0.18$ & --- & $-0.29\pm0.18$&  $1.2\times10^{23}$ & $55.8\pm0.1$ \\ 
1456+044 & 0.391 & 20.15 & 5.57 &$19.55\pm1.20$ &  $4.64\pm0.28$   & $1.22\pm0.07$ & --- & $-0.52\pm0.07$ &  $2.0\times10^{21}$& $54.2\pm0.6$ \\ 
1509+022 & 0.219 & 19.83 & 5.41 & $18.81\pm0.35$ & $5.16\pm0.10$ & $1.03\pm0.14$&  --- & $-0.65\pm0.14$ & $9.1\times10^{20}$  & $53.8\pm0.1$\\ 
\hline
\end{tabular}
\label{targets}  
\end{minipage}
\end{table*} 

Verifying that the magnitude selection did yield targets exceeding the critical UV luminosity/ionising photon rate, 
\begin{figure} 
\centering \includegraphics[angle=-90,scale=0.58]{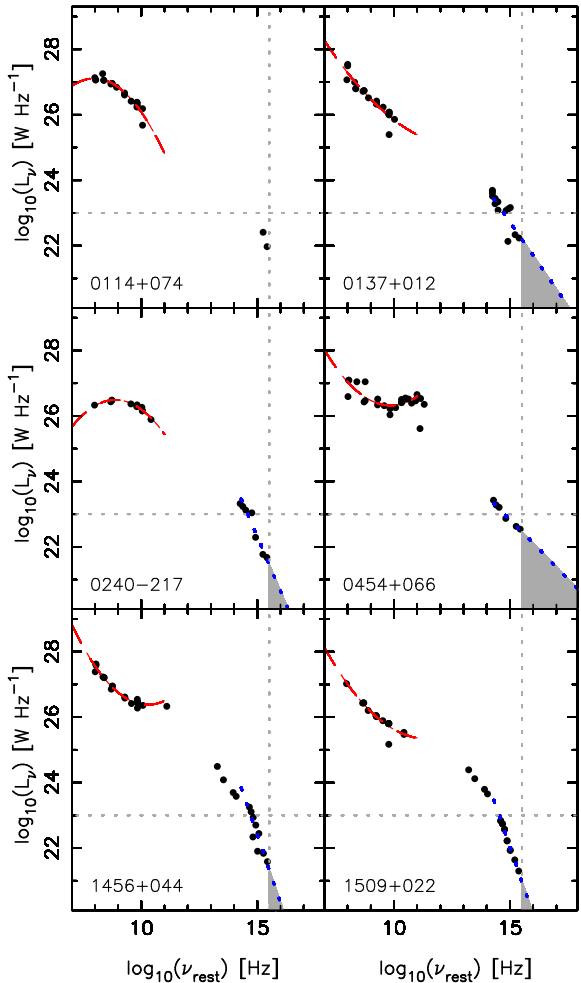} 
\caption{The rest-frame spectral energy distribution (SED) for each of  our targets overlaid by fits to the photometry.
The broken curve shows the second order polynomial fit to the radio data 
and the dotted line the power-law fit to the UV data.
The vertical dotted line signifies a rest-frame frequency of $3.29\times10^{15}$ Hz ($\lambda = 912$~\AA) 
and the horizontal the critical $\lambda = 912$~\AA\  luminosity of $L_{\rm UV}\sim10^{23}$ \WpHz, with the shading
showing the region over which the ionising photon rate is derived.}
\label{SEDs}
\end{figure}
as described in \citet{cwsb12}, for each source we obtained the photometry from NASA/IPAC Extragalactic Database (NED),
the Wide-Field Infrared Survey Explorer (WISE, \citealt{wem+10}), Two Micron All Sky Survey (2MASS, \citealt{scs+06})
and the Galaxy Evolution Explorer (GALEX, data release GR6/7)\footnote{http://galex.stsci.edu/GR6/\#mission} databases. Each datum was corrected for Galactic extinction
\citep{sfd98} and a power law was fit to the
UV rest-frame data, allowing the ionising photon rate, $Q_\text{\HI}\equiv \int^{\infty}_{\nu}({L_{\nu}}/{h\nu})\,d{\nu}$, to be derived from
\[
\int^{\infty}_{\nu}\frac{L_{\nu}}{h\nu}\,d{\nu},~{\rm where}~\log_{10}L_{\nu} = \alpha\log_{10}\nu+ {\cal C} \Rightarrow  L_{\nu} = 10^{\cal C}\nu^{\alpha},
\]
where \AL\ is the spectral index and ${\cal C}$ the intercept. This gives,
\[
\frac{10^{\cal C}}{h}\int^{\infty}_{\nu}\nu^{\alpha-1}\,d{\nu} = \frac{10^{\cal C}}{\alpha h}\left[\nu^{\alpha}\right]^{\infty}_{\nu} = \frac{-10^{\cal C}}{\alpha h}\nu^{\alpha}~{\rm where}~\alpha < 0,
\]
shown by the shaded regions in Fig. \ref{SEDs}, all of which give rates below the critical $\log_{10}Q_\text{\HI}=56.5$~sec$^{-1}$  (Table \ref{targets}).

\begin{figure}
\centering \includegraphics[angle=-90,scale=0.6]{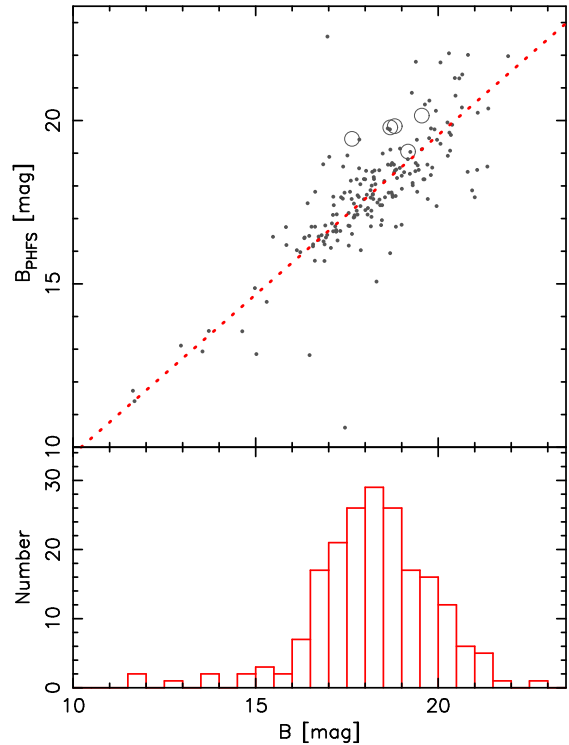}
\caption{The PHFS blue magnitudes (from COSMOS) versus those derived from our SED fits, where possible. The large
  circles designate our targets and the line the least-squares fit. The gradient of this is 0.977 (regression
  coefficient 99.46), showing that our $B$ estimates are, on average, slightly higher than those in the PHFS.}
\label{B-histo}
\end{figure}

Fitting the optical photometry by the same method gave the blue and near-infrared magnitudes, where the latter would be
primarily obtained directly from the WISE and 2MASS data. From this we found brighter blue magnitudes than
expected. \citet{dwf+97} obtain the blue magnitudes for the PHFS from the COSMOS catalogue \citep{ycg+92}, where the
uncertainty quoted is $\pm0.5$ mag, although \citeauthor{dwf+97} find that, due to a lack of calibration, some
magnitudes may be erroneous by $>1$ mag. In order to check this, in Fig. \ref{B-histo} we compare the blue magnitudes
obtained from the fit to the SED of each PHFS source with those quoted in the catalogue.
From a least-squares fit to the distribution, we see that, although our fits suggest larger $B$ values ($\mu = 18.19\pm0.12$ for the
whole sample) than quoted in the PHFS ($\mu = 17.80\pm0.14$), the magnitudes of our targets may be consistent with the PHFS values
within the uncertainties (Table \ref{targets}). However, the large possible uncertainties in the magnitudes were 
 not a consideration when selecting targets and so these sight-lines are generally not be as reddened as originally thought.

\subsection{Observations and data reduction}
\label{sec:obs}

All of the targets were observed over 26--31 January 2012, with the 1420 MHz receiver backed with the FX correlator over
a bandwidth of 16 MHz spread over 512 channels in orthogonal polarisations. This gave a channel spacing of $\approx8$
\kms\ (cf. a full-width half maximum of FWHM = 8 to 210 \kms\ for the five known OH absorbers, see \citealt{cdbw07}),
while maintaining a redshift coverage of $\Delta z \approx \pm0.01$, in order to cover any uncertainties in the
redshifts.\footnote{These are $\delta z = 0.000011$ for 0137+012 \citep{hk09}, $\delta z = 0.00035$ for 1456+044 and
  $\delta z = 0.00035$ for 1509+022 \citep{hw10}. Uncertainties are not available for the remaining three targets, but since the 
redshifts are quoted to three decimal places, the choice of $\Delta z \approx \pm0.01$ should ensure ample coverage.}
The full 30 antenna array
was requested, with each transition in each source being observed for a total of one hour, in order to reach a root mean-square
(rms) noise level of $\approx1$ mJy per channel. For a flux density of $S_{\rm fit}\gapp0.5$~ Jy,
this gives a $3\sigma$ optical depth limit of $\tau\approx0.005$ per channel, or a sensitivity to $N_{\rm
  HI}\sim1\times10^{17}.\,(T_{\rm spin}/f)$ \scm\ per 8 \kms\ channel, which is close to the lower limit for all of the
published \HI\ 21-cm searches.  For the OH 1667 MHz transition, this corresponds to a sensitivity to $N_{\rm
  OH}\sim2\times10^{13}.\,(T_{\rm ex}/f)$ \scm\ per 8 \kms\ channel, cf. $0.55 - 56\times10^{13}.\,(T_{\rm ex}/f)$ \scm\
for the five known OH absorbers \citep{kc02a,kcl+05}.

For each source, 3C\,48, 3C\,147 and 3C\,286 were used for bandpass calibration and a strong nearby point source for
phase calibration.  However, since this was performed only once every 30 minutes, self calibration of the delays usually
produced a superior image. The data were flagged and reduced using the {\sc miriad} interferometry reduction package, with
flagging of the edge channels leaving the central 470 channels ($\approx\pm2000$ \kms) from which a bad channel (144 in
polarisation XX) was removed from all visibilities. After averaging the two polarisations, a spectrum was extracted from the cube. 
Regarding each source:

{\bf 4C\,+07.04 (0114+074)}: The \HI\ band was observed for a total of 1.05 hours. 
Five antennas (14, 20, 22, 23 \& 30) were non-functioning, with antennas 19 and 26 also being flagged due to a severe
bandpass ripple, leaving 300 baseline pairs. Self calibration of the phases produced a superior image in which the source 
was unresolved by the $5.6\arcsec\times4.3\arcsec$ synthesised beam. \\
The OH band was observed for a total of 1.27 hours and after flagging of the non-functioning antennas (14,20,22,23,30),
300 baseline pairs remained. Again, a far superior image was obtained through self calibration of the phases, giving a
synthesised beam of $4.1\arcsec\times3.5\arcsec$.  The extracted spectrum shows a strong emission feature, which we
believe to be an artifact (see Sect. \ref{sec:ef}).
\\
{\bf UM\,355 (0137+012)}: \HI\ was searched for a total of 1.05 hours.  After flagging non-functioning antenna 20,
antennas 23, 28, 29 and 30 were removed, due to less than ideal phase calibration. Although all of the
remaining phases were well behaved, self calibration of this source could not produce a satisfactory image. The
calibration was then obtained from the phase calibrator LBQS\,0056--0009, and the data from the unflagged 351
baseline pairs. The synthesised beam was $5.3\arcsec\times3.4\arcsec$ giving the partially resolved main component and a separate
feature (see Sect. \ref{sec:rs}).  The OH band was observed for a total of 1.28 hours. After removal of the non-functioning
antennas (12, 20, 22, 23 and 30), 300 good baseline pairs remained. \\

{\bf PKS\,0240--217}: The \HI\ band was observed for a total of 0.90 hours.  Only one antenna (20) was non-functioning,
leaving 406 baseline pairs. Some minimal RFI was removed from the first half of the observation and after self
calibration of the phases a high quality image was produced, although there is still some narrow-band RFI spikes present
in the extracted spectrum. Unfortunately, this is concentrated at $\approx1082$ MHz, close to the expected absorption
frequency (Fig. \ref{spectra}).  The source was unresolved by the $10.4\arcsec\times3.4\arcsec$ synthesised
beam.\\ 
For the OH band, non-functioning antennas (4 and 20) were flagged, leaving 378 baseline pairs.  Again, by self
calibrating, an excellent image was produced from which the spectrum was extracted. The source was unresolved by the
$4.7\arcsec\times2.6\arcsec$ synthesised beam.\\ 
{\bf 4C\,+06.21 (0454+066)}: \HI\ was searched for a total of 0.66 hours.  In addition to antenna 20, antenna 4 was
found to be non-functioning with antenna 26 giving extremely noisy spectra. After flagging these, 351 baselines pairs
remained from which the image was produced, using 3C\,120 for phase calibration. The source was unresolved by the
$4.3\arcsec\times3.2\arcsec$ synthesised beam.\\ 
For the OH band, antenna 20 was non-functioning and after the removal of the badly behaving antenna 12, 378 baseline
pairs remained. However, RFI was apparent below 1182 MHz requiring the first 200 channels to be flagged, leaving 290.
The source was unresolved by the $3.3\arcsec\times2.7\arcsec$ synthesised beam.\\ 
{\bf 4C\,+04.49 (1456+044)}: \HI\ was searched for a total of 0.47 hours.  After removing non-functioning antennas (4
and 26), the non-functioning baseline pair 2--8 was also removed, leaving 377 pairs. Self calibration of the phases
proved unsatisfactory and so the nearby 3C\,298 (1416+067) was used. Severe RFI below 1016 MHz required removal of the
first 70 remaining channels, leaving 420.
The $4.9\arcsec\times3.4\arcsec$ beam reveals a double source (see Sect. \ref{sec:rs}). \\
For the OH band, non-functioning antennas (4 and 26) were removed, leaving 378 baseline pairs. RFI was present on all
baselines at $\gapp1200$ MHz and using 3C\,298 for the phase calibration also revealed the second feature present in the
lower frequency, resolved by the $3.2\times2.9\arcsec$ beam. \\ 
{\bf PKS\, 1509+022}: Was searched in the \HI\ band for a total of 0.73 hours. After flagging the non-functioning
antennas (4, 22, 23 and 26), phase calibration, using 3C\,327.1, failed to produce a quality image even, after the
removal of poorly performing antennas (17, 24, 28 and 30), which left 231 baseline pairs.  This was probably due to the
fact that this calibrator is over one hour in distance from the target source. Self calibration of the source required
removal of the aforementioned antennas before the phases could be calibrated. Even so, no good image could be produced
and so the spectrum was obtained by averaging visibilities of the remaining 231 baselines.\\ 
The OH band was observed for a total of 0.85 hours and, like the \HI\ band observations, could not be calibrated, even
after the removal of non-functioning antennas. 
Further flagging, of the noisiest baseline pairs (where the rms exceeded 1 Jy), left 210 baseline pairs, which were
averaged together to obtain a spectrum.

\section{Results}
\label{sec:res}
\subsection{Observational results}
\label{sec:os}

\begin{figure*}
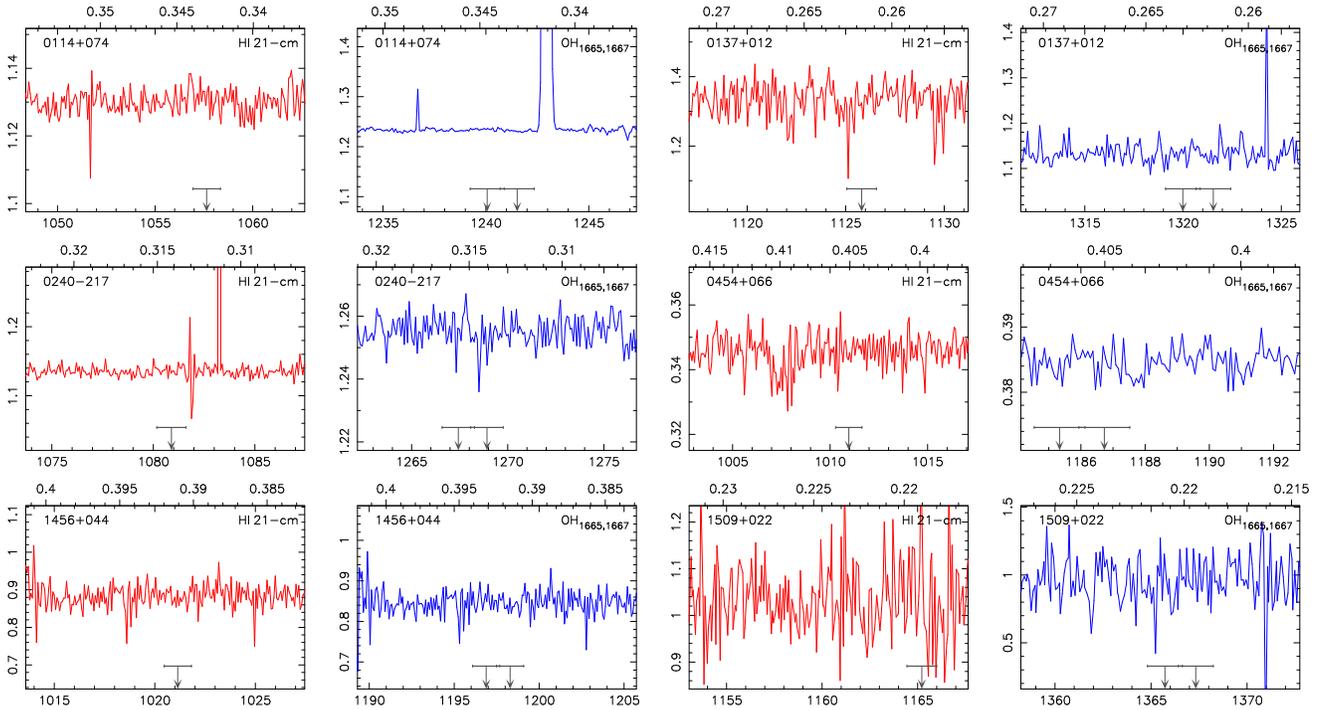
 
\vspace{9.3cm}  
\includegraphics{spectra/0114+074_HI.dat-freq_poly2_flux_20kms.ps}  
\includegraphics{spectra/0114+074_OH.dat-freq_poly2_flux_20kms.ps} 
\includegraphics{spectra/0137_HI-all.dat-freq_poly3_flux_20kms.ps} 
\includegraphics{spectra/0137_OH-all.dat-freq_poly5_flux_20kms.ps}
\includegraphics{spectra/0240-217_HI_self.dat-amp-freq_poly1_flux_20kms.ps} 
\includegraphics{spectra/0240-217_OH_self.dat-amp-freq_poly1_flux_20kms.ps}  
\includegraphics{spectra/0454+066_HI.dat-freq_poly1_flux_20kms.ps} 
\includegraphics{spectra/0454+066_OH.dat-freq_poly2_flux_20kms.ps} 
\includegraphics{spectra/1456-HI_all.dat-freq_poly2_flux_20kms.ps}  
\includegraphics{spectra/1456+044_OH-all.dat-freq_poly2_flux_20kms.ps} 
\includegraphics{spectra/1509+022_HI_phase_uv_231.dat-amp-freq_poly5_flux_20kms.ps} 
\includegraphics{spectra/1509+022.1365_OH_uv.dat-amp-freq_poly1_flux_20kms.ps} 
 \caption{The spectra from the \HI\ 21-cm and OH 18-cm main line (1665 \& 1667 MHz)  searches shown at  a spectral resolution of 20 \kms. 
 All are extracted from the cube apart from the 1509+022 spectra, which are  averages of the unflagged visibilities.
The ordinate gives the flux density [Jy] and the abscissa the barycentric frequency [MHz]. The scale along the top shows the redshift
of either the \HI\ 1420 MHz or OH 1667 MHz transition over the frequency range  and the downwards arrow shows the expected
frequency of the absorption from the optical redshift, with the horizontal bar showing a span of $\pm200$ \kms\ for guidance 
(the profile widths of the \HI\ 21-cm detections range from 18 to 475 \kms, with a mean of 167 \kms, \citealt{cwsb12}).}
\label{spectra}
\end{figure*}

 In Fig. \ref{spectra} we show the final spectra from which we have obtained no detections in the six targets searched, 
with the details summarised in Table \ref{obs}.
\begin{table*} 
\centering
\begin{minipage}{165mm}
  \caption{The observational results. $z$ is the optical redshift of the source, $S_{\rm
      meas}$ is the measured flux density, $\Delta S$ the rms noise reached per $\Delta v$ channel
    and $\tau$ the derived optical depth, where $\tau_{_{3\sigma}}=-\ln(1-3\Delta S/S_{\rm cont})$ is quoted for these
    non-detections.  These give the quoted column densities [\scm], where $T_{\rm s}$ is the spin temperature of the
    atomic gas, $T_{\rm x}$ is the excitation temperature [K] of the molecular gas and $f$ the respective covering factor.  Here,
    and throughout the paper, OH refers to the $^{2}\Pi_{3/2} J = 3/2$ $F=2-2$ (1667 MHz) transition.  Finally, we list
    the frequency and redshift range over which the limit applies.}
\begin{tabular}{@{}l l   c  c c c  r c c  c  @{}} 
\hline
\smallskip
Source &   $z$  & Line & $S_{\rm meas}$   [Jy] &$\Delta S$ [Jy] & $\Delta v$ [\kms]& $\tau_{_{3\sigma}}$ & $N$ [\scm] & $\nu$-range [MHz] & $z$-range \\
\hline
0114+074  & 0.343 & \HI & 1.130 & 0.003   & 9.2 & $<0.009$ & $<1.5\times10^{17}\,(T_{\rm s}/f)$ & 1048--1063& 0.336--0.356 \\
...        & ...     &  OH   &  1.233& 0.004   &  7.9  &  $<0.010$ &  $<1.9\times10^{13}\,(T_{\rm x}/f)$ &  1232--1247  &     0.337--0.354    \\ 
0137+012   & 0.260 & \HI & 1.335 &    0.052 & 8.7 & $<0.11$ &     $<1.8\times10^{18}\,(T_{\rm s}/f)$    & 1119--1134 &    0.252--0.269\\
... & ...      & OH  &  1.133 & 0.022 & 7.4 & $<0.058 $  &  $<1.0\times10^{14}\,(T_{\rm x}/f)$    & 1315--1327 &   0.256--0.268\\ 
0240--217 & 0.314 & \HI & 1.135 & 0.007 &  9.0 & $<0.019$  &  $<3.0\times10^{17}\,(T_{\rm s}/f)$ &  1073--1088&    0.306--0.324 \\ 
 ... & .. & OH  &   1.255 & 0.005 & 7.7 & $<0.012$&   $<2.2\times10^{13}\,(T_{\rm x}/f)$   & 1262--1277 &      0.306--0.321 \\ 
0454+066 & 0.405 & \HI & 0.346 & 0.005 & 9.7 & $<0.041$&  $<7.2\times10^{17}\,(T_{\rm s}/f)$  & 1002--1017& 0.396--0.417 \\ 
...              & ...& OH  & 0.385& 0.002 & 8.2 & $<0.016$&   $<3.1\times10^{13}\,(T_{\rm x}/f)$  & 1184--1193&    0.398--0.409\\
 1456+044 & 0.391 & \HI &  0.884& 0.033& 9.6 & $<0.11$& $<2.0\times10^{18}\,(T_{\rm s}/f)$  & 1015--1028& 0.382--0.400\\ 
...                 & ... &  OH  &  0.846 & 0.033 & 8.1 & $<0.12$ &    $<2.3\times10^{14}\,(T_{\rm x}/f)$ & 1193--1202& 0.387--0.397  \\ 
1509+022 & 0.219 & \HI\ & 1.034 & 0.08 & 8.5  & $<0.23$  & $<3.6\times10^{18}\,(T_{\rm s}/f)$   & 1153--1168  & 0.216--0.232 \\
 ...    & ...  & OH  &  0.972 & 0.16 & 7.1 &    $<0.68$& $<1.2\times10^{15}\,(T_{\rm x}/f)$  & 1358--1373&  0.214--0.228\\
\hline
\end{tabular}
\label{obs}  
\end{minipage}
\end{table*} 
In addition to our targets not being sufficiently red (Sect.~\ref{ss}) to detect OH \citep{cwm+06}, 
each of the five known redshifted systems OH  was detected following a clear \HI\ detection
\citep{cps92,cry93,cmr+98,cdn99,kb03}. Furthermore,
the OH absorption strength is only expected to be $\lapp10^{-4}$ times that of the \HI\ 21-cm absorption
\citep{cdbw07}. Therefore we will treat \HI\ as a prerequisite for OH absorption and thus focus our discussion around
the \HI\ 21-cm results.

\begin{figure*}
\centering \includegraphics[angle=-90,scale=0.66]{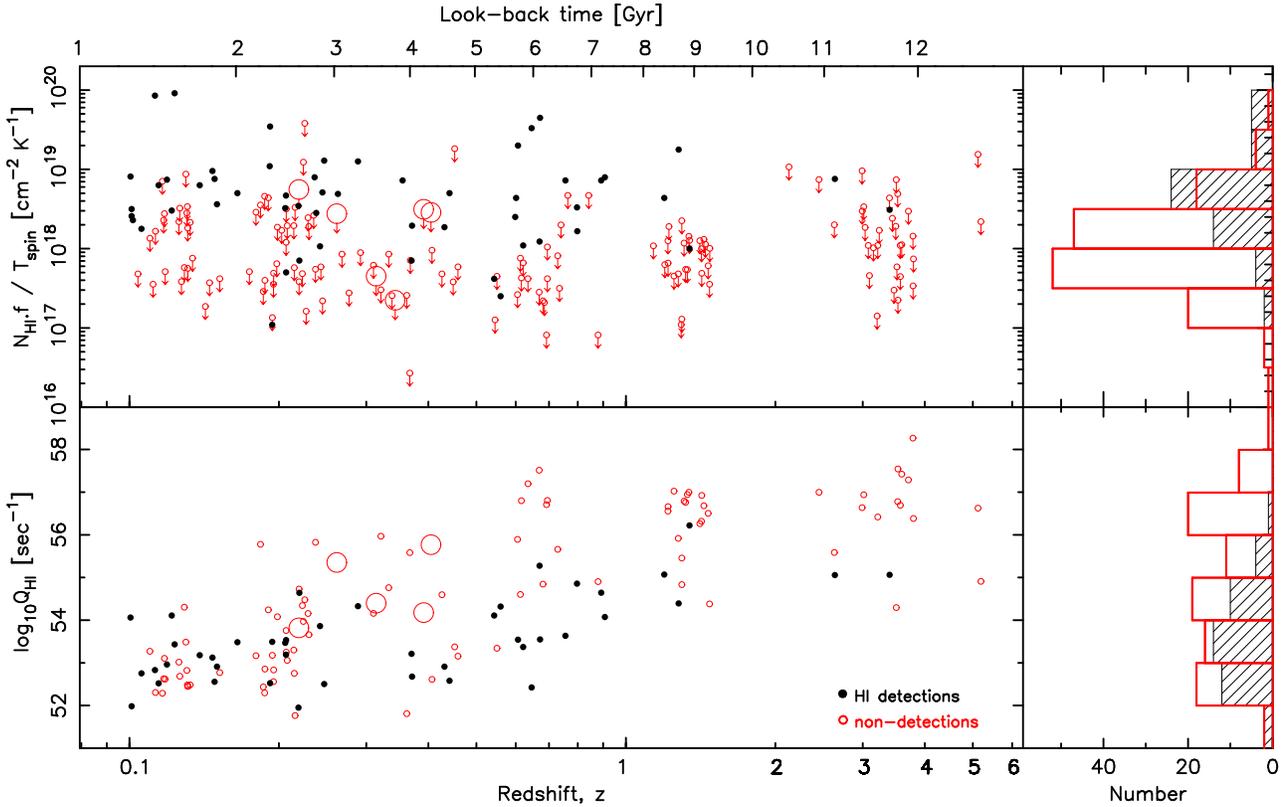}
\caption{The line strength ($1.823\times10^{18}\,(T_{\rm  spin}/f) \int\!\tau_{\rm obs}\,dv$, top) and 
ionising ($\lambda \leq912$ \AA) photon rate (bottom) versus redshift for the $z\geq0.1$ \HI\ 21-cm absorption searches. The filled circles/histogram
  represent the detections and the unfilled circles/histogram the $3\sigma$ upper limits to the non-detections, with the
  large circles designating our targets.}
\label{N-ion}
\end{figure*}

The total neutral hydrogen column density is related to the velocity integrated optical depth of the \HI\ 21-cm absorption via,
\begin{equation}
N_{\text \HI}  =1.823\times10^{18}\,T_{\rm  spin}\int\!\tau\,dv,
\label{enew_full}
\end{equation}
where $T_{\rm spin}$ is the spin temperature of the gas, which is a measure of the excitation from the lower hyperfine
level \citep{pf56,fie59}, and $\int\!\tau dv$ is the velocity integrated optical depth of the absorption. The observed
optical depth is related to this via
\begin{equation}
\tau \equiv-\ln\left(1-\frac{\tau_{\rm obs}}{f}\right) \approx  \frac{\tau_{\rm obs}}{f}, {\rm ~for~}  \tau_{\rm obs}\equiv\frac{\Delta S}{S_{\rm obs}}\lapp0.3,
\label{tau_obs}
\end{equation}
where the covering factor, $f$, is a measure of the fraction of observed background flux ($S_{\rm obs}$) intercepted by
the absorber.  In the optically thin regime (where $\tau_{\rm obs}\lapp0.3$), Equ. \ref{enew_full} can be rewritten as
$N_{\text \HI}\approx 1.823\times10^{18}\,(T_{\rm spin}/f)\int\!\tau_{\rm obs}\,dv$.  From Fig. \ref{N-ion} (top
panel)\footnote{In order to normalise the limits, each has been re-sampled to the same spectral resolution
  (20 \kms, as in Fig. \ref{spectra}), which is used as the FWHM to obtain the integrated optical depth limit, thus giving the 
$N_{\text \HI} f/T_{\rm spin}$ limit per channel (see \citealt{cur12}).}, 
we see that the six targets have been searched in \HI\ 21-cm to sensitivities comparable with previous detections.\footnote{The associated
absorption searches have been compiled from \citet{dsm85,mir89,vke+89,ubc91,cps92,cmr+98,cwh+07,mcm98,ptc99,ptf+00,rdpm99,mot+01,ida03,vpt+03,cwm+06,cww+08,cwm+10,cwwa11,cwsb12,cwt+12,caw+16,gss+06,omd06,kpec09,ems+10,ssm+10,css11,cgs13,ace+12,asm+15,ysdh12,ysd+16,gmmo14,sgmv15,akk16,akp+17}.}

\begin{figure}
\centering 
\includegraphics[angle=270,scale=0.52]{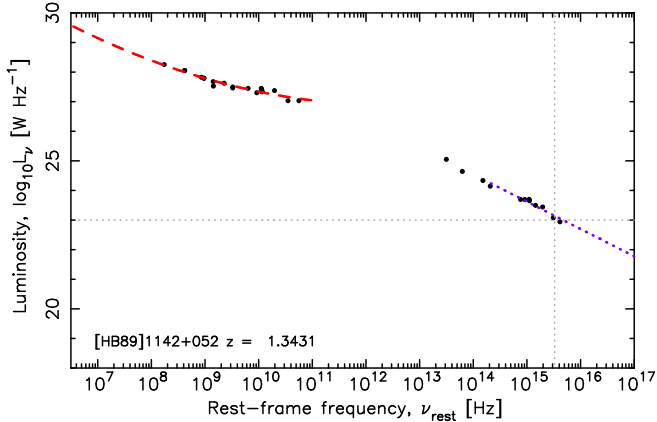}  
\caption{The rest-frame SED of [HB89]\,1142+052 overlaid by fits to the radio and optical--UV photometry (from the
Sloan Digital Sky Survey data releases 2 \& 6, \citealt{aaa+04,aaa+08}). As per
  Fig. \ref{SEDs}, the vertical dotted line signifies a rest-frame frequency of $3.29\times10^{15}$ Hz ($\lambda =
  912$~\AA) and the horizontal the critical $\lambda = 912$~\AA\ luminosity of $L_{\rm UV}\sim10^{23}$ \WpHz.}
\label{1142}
\end{figure}

In the bottom panel we show the ionising photon rates for those sources for which there is sufficient blue/UV rest-frame
photometry (Sect. \ref{ss}). [HB89]\,1142+052 now defines the highest value ($Q_\text{\HI}=1.66_{-0.49}^{+0.70}\times10^{56}$~sec$^{-1}$)
where \HI\ has been detected \citep{kpec09}.\footnote{The absorption redshift of $z_{\rm abs} = 1.343$ \citep{kpec09}
  corresponds to a velocity offset of $\Delta v = 64$ \kms\ relative to $z_{\rm QSO} = 1.3425$ and so we deem this as
  associated rather than intervening.}  This is close to the theoretical value of $ 3\times10^{56}$~sec$^{-1}$, which is
sufficient to ionise all of the neutral atomic gas in a large spiral galaxy \citep{cw12}, although from very limited
$\lambda > 912$~\AA\ photometry (Fig. \ref{1142}).
For $Q_\text{\HI}\leq1.7\times10^{56}$~sec$^{-1}$, there are 43 detections and 66 non-detections, that is a 39.4\%
detection rate for objects for which $Q_\text{\HI}$ can be estimated.  Applying this to the $Q_\text{\HI}>
1.7\times10^{56}$~sec$^{-1}$ sources, gives a binomial probability of $4.80\times10^{-7}$ of obtaining 0 detections and
29 non-detections, which is significant at $5.03\sigma$, assuming Gaussian statistics.  If we include the forthcoming
results of \citet{gdb+15}\footnote{Hereafter G17, which reports 0 new detections of \HI\ 21-cm absorption out of 89 new
  searches over $0.02 < z < 3.8$ (see \citealt{gd11}).}, this increases to $6.67\sigma$.

\citet{akp+17} have recently detected \HI\ 21-cm absorption at $z = 1.223$ in TXS\,1954+513, which they claim has an
ultra-violet luminosity of $L_{\rm UV}= 4\pm1\times10^{23}$ \WpHz, following the method of \citet{cww+08}. The
luminosity is, however, based upon only two photometry measurements which are in the rest-frame optical band and
extrapolated to the UV. This is insufficient data to calculate $Q_\text{\HI} \equiv
\int^{\infty}_{\nu}({L_{\nu}}/{h\nu})\,d{\nu}$, which requires integration of the UV photometry above $\nu_{\rm rest} =
3.29\times10^{15}$ Hz (see Sect. \ref{ss}).  If, for the sake of argument, we approximate $Q_\text{\HI} \sim L_{\rm
  UV}/h\sim6\times10^{56}$~sec$^{-1}$, our model \citep{cw12} yields a scale-length of $R=3.05$ kpc, cf.  $3.15$ kpc for
the Milky Way \citep{kk09}.  That is, this ionising photon rate is just sufficient to ionise all of the gas in a large
spiral, which is consistent with the results of \citeauthor{cw12}. Adding this value of $Q_\text{\HI}$ to the other data
(Fig. \ref{N-ion}, bottom), would change the above significance to $3.30\sigma$ for the current data and $4.50\sigma$
including the forthcoming data. Due to the lack of photometry, however, our own UV fitting method (which requires the
interpolation of at least four points at $\nu_{\rm rest} > 4\times10^{14}$ Hz, see \citealt{cwsb12}) would reject a UV
fit for this source and so we cannot assign an ultra-violet luminosity nor photo-ionisation rate.

\subsection{Emission feature in 0114+074}
\label{sec:ef}

\begin{figure} 
\centering 
\includegraphics[angle=270,scale=0.45]{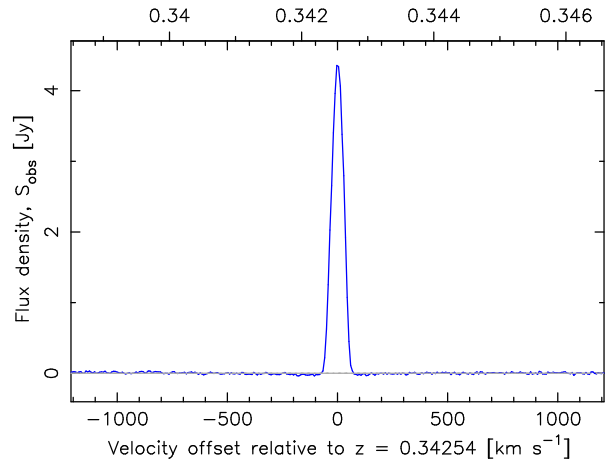}
\caption{  Detail of the ``emission'' feature towards 0114+074 shown at the observed $\Delta v = 7.85$ \kms\  
resolution (cf. Fig.~\ref{spectra}).
The scale along the top shows the redshift of the OH 1667 MHz transition, which is also used to define
the velocity offset.}
\label{off}
\end{figure}

As seen in Fig. \ref{spectra}, there is a strong emission feature in the OH band of 0114+074. The feature was apparent
in the averaged visibilities, before imaging, in both polarisations, close to the expected OH frequency
(Fig. \ref{off}), although not apparent in the calibration sources.
The emission has an integrated flux density of $\int S_{\rm obs} dv = 286\pm9$ Jy \kms, which gives $L_{\rm OH} = 870$ \Lo. This is  
within the range of known OH mega-masers ($10^{2.2} < L_{\rm OH} < 10^{3.8}$~\Lo, e.g. \citealt{dg02a,lo05}). However, given that the 
feature appears as absorption when off source, in conjunction with the fact that no 1665/1667 MHz doublet is apparent
(e.g. \citealt{dg02,hag+16}), we  attribute this to an artifact rather than being the detection of an OH mega-maser.

\subsection{Resolved structure in 0137+012 \& 1456+044}
\label{sec:rs}
			 
\begin{figure} 
\vspace{11.9cm}  
\includegraphics{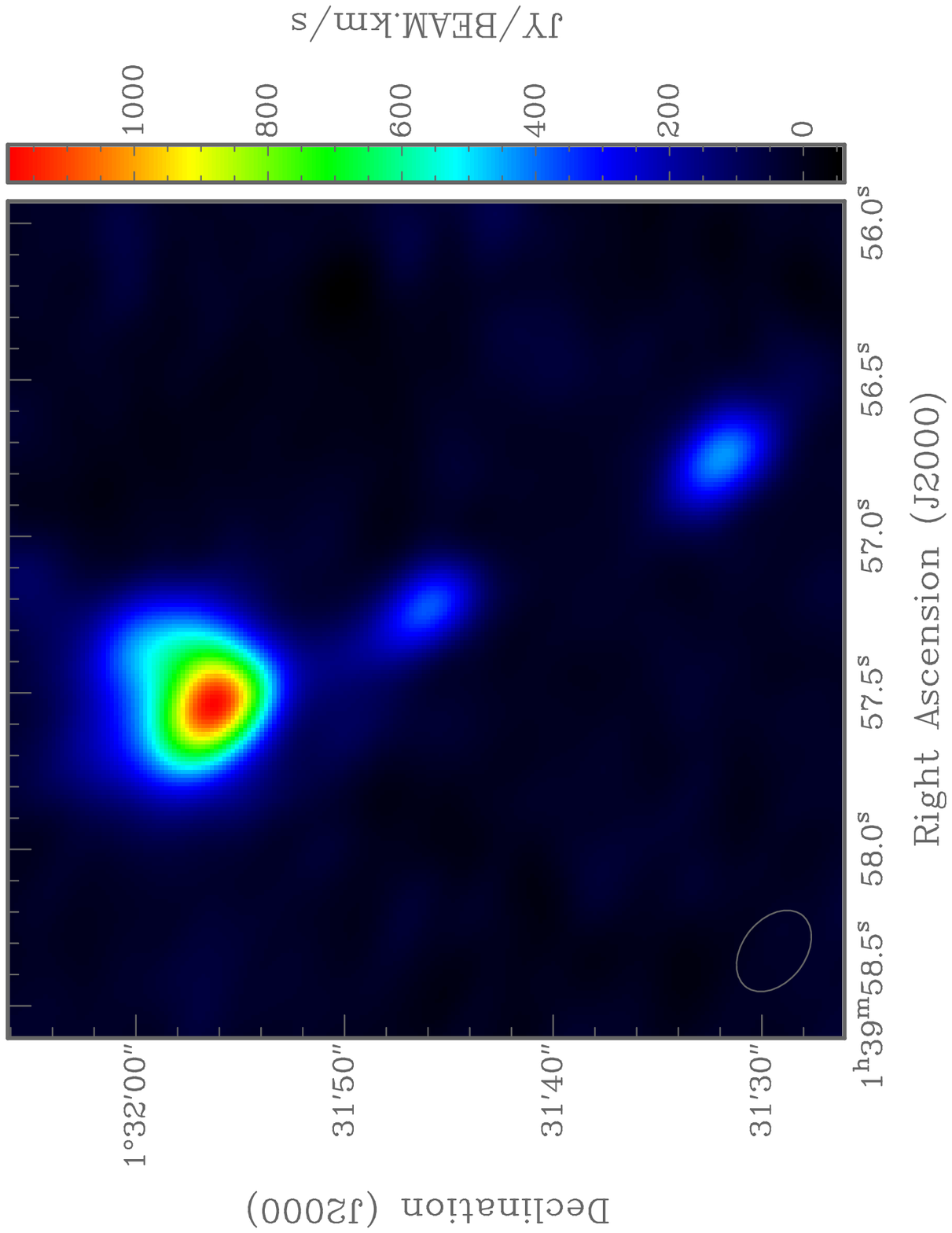}  
\includegraphics{0137_full.ps} 
\caption{Top: GMRT 1126 MHz continuum image of 0137+012. The $5.3\arcsec\times3.4\arcsec$ synthesised beam is shown in
  the bottom left corner. Bottom: The spectral indices of the various components (designated 1--3 starting from the
  north) obtained from the redshifted \HI\ and OH band observations. The small circles show the data from
the literature overlain with the polynomial fit (see Fig. \ref{SEDs}).}
\label{0137}
\end{figure}

As mentioned in Sect.~\ref{sec:obs}, more than one component was resolved in the observations of 0137+012 and 1456+044.
In 0137+012  (Fig.~\ref{0137}, top) we see three separate features, previously detected at 1465 and 4885 MHz
by \citet{gh84a}, which they modelled as a damped precessing jet.
Component 1 is by far the strongest, with flux densities of $0.801\pm0.008$ and $0.636\pm0.005$~Jy at 1125 and 1321 MHz,
respectively, giving an \HI\ optical of depth limit of $\tau_{3\sigma} < 0.029$ per 8.7~\kms\ channel. The other two
components are significantly weaker (2 -- $0.125\pm0.006$~Jy \& $0.122\pm0.008$~Jy, giving $\tau_{3\sigma} < 0.14$ and 3 --
$0.155\pm0.005$~Jy \& $0.150\pm0.003$~Jy, giving $\tau_{3\sigma} < 0.10$). They also have similar spectral indices which  may suggest a
connection, although the two frequencies are insufficiently separated to infer anything definite (Fig.~\ref{0137}, bottom).

\begin{figure} 
\vspace{11.9cm}  
\includegraphics{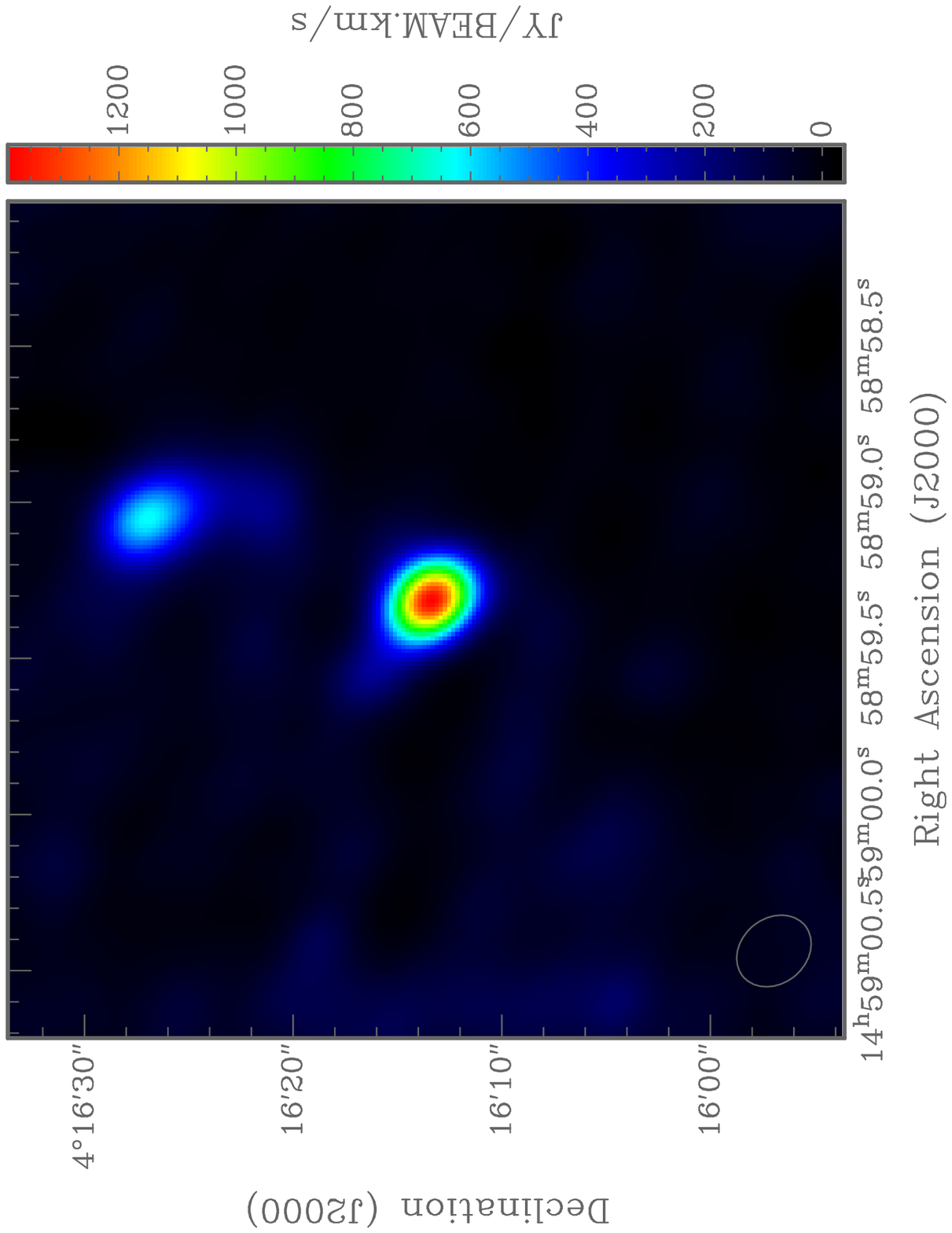}  
\includegraphics{1456_full.ps} 
\caption{Top: GMRT 1021 MHz continuum image of 1456+044. The $4.9\arcsec\times3.4\arcsec$ synthesised beam is shown in the bottom
  left corner. Bottom: The spectral indices of the various components, designated 1 (main) and 2 (north), obtained from
  the redshifted \HI\ and OH band observations.}
\label{1456}
\end{figure}

For 1456+044, the image revealed a  double source  (Fig. \ref{1456}, top).
Component 1 has flux densities of $0.371\pm0.010$ and $0.365\pm0.09$ Jy at 1021 and 1198 MHz, respectively, giving an
\HI\ optical of depth limit $\tau_{3\sigma} < 0.081$ per 8.5 \kms\ channel. For component~2, the flux densities are
$0.207\pm0.06$ and $0.180\pm0.06$~ Jy, giving an \HI\ optical of depth limit $\tau_{3\sigma} < 0.093$.  From
Fig. \ref{1456} (bottom), while the main component appears to have a very flat spectrum, component 2 has a similar
spectral index to the cumulative emission at the measured frequencies. Furthermore, it is possible to envision the
cumulative SED as being the sum of those of the two components, although there appears to be some missing flux and, again,
the observed frequencies are very close to each other.

\section{Discussion}
\label{sec:disc}
\subsection{Factors affecting the detection rate}
\subsubsection{Ultra-violet luminosity and survey sensitivity}

Although all six sources searched have UV luminosities below the critical value, none were detected in \HI\ 21-cm
absorption.  
\begin{figure*}
\centering 
\includegraphics[angle=270,scale=0.63]{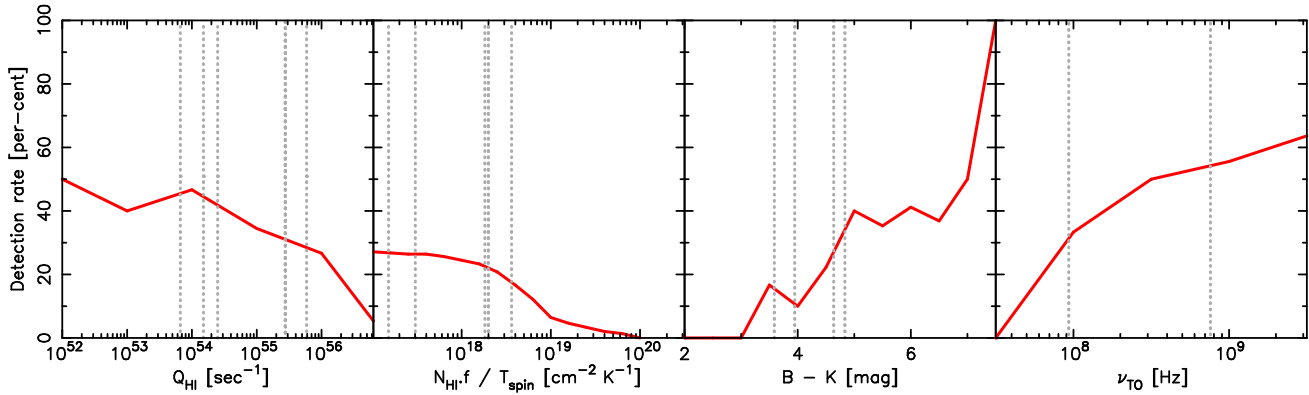}
\caption{The \HI\ 21-cm detection rate binned against the ionising photon rate, the sensitivity, the blue -- near-infrared colour and
  the turnover frequency (clear turnovers over the observed range only, see Fig. \ref{SEDs}). These are binned to 
steps of (log) unity on the abscissa and intended only to show the overall trends, thus no uncertainties are shown.
The vertical dotted lines show the values for our sample, where two overlap in the $N_{\rm HI}f/T_{\rm spin}$ and $B-K$
panels.}
\label{all-rate}
\end{figure*} 
In Fig.~\ref{all-rate}, for the photo-ionising rate and other important parameters, we show how the detection
rate varies with the parameter in question.  These are obtained from the number of detections normalised by the total number of
searches within each bin, which we bin per decade for $Q_\text{\HI}$ (the first panel of Fig.~\ref{all-rate}). For our
targets, whose values are shown by the vertical lines, we may expect a $28-52$\% chance of detection. A decrease in the
detection rate with redshift was first shown by \citet{cww+08}, which was interpreted as the high redshifts selecting UV
luminosities sufficient to excite the gas to below the detection limit.\footnote{\citet{rdpm99} also obtained only \HI\
  21-cm non-detections in six $z\gapp2.4$ radio galaxies, but failed to realise the significance of the UV luminosity in
  their comparison with the highest redshift ($z=3.40$) \HI\ 21-cm detection \citep{ubc91}. This, B2\,0902+34, has
  $L_{\rm UV}\approx3\times10^{22}$~\WpHz, cf. $L_{\rm UV}\gapp10^{23}$ \WpHz\ for each of their sample.}  Because of
the Malmquist bias, it can be difficult to ascertain whether the decreasing detection rate is caused by an evolutionary
effect at high redshift \citep{cw10,akk16}, although ionisation of the gas at high UV luminosities is physically
motivated 
and is found to apply at {\em all} redshifts, no matter the selection criteria (Sect.~\ref{intro}).  Showing the line
strength versus the photo-ionising rate (Fig. \ref{Q-N}), we see a strong anti-correlation, which, for a given column
density and covering factor (Sect. \ref{sec:os}), would suggest an increase in the spin temperature due to an
increased flux of $\lambda\leq912$~\AA\ photons \citep{be69}. Again, it is also clear that there is a value of $Q_\text{\HI}$
above which \HI\ 21-cm is not detected, although there is no ``critical redshift'' apparent. 
\begin{figure}
\centering 
\includegraphics[angle=270,scale=0.52]{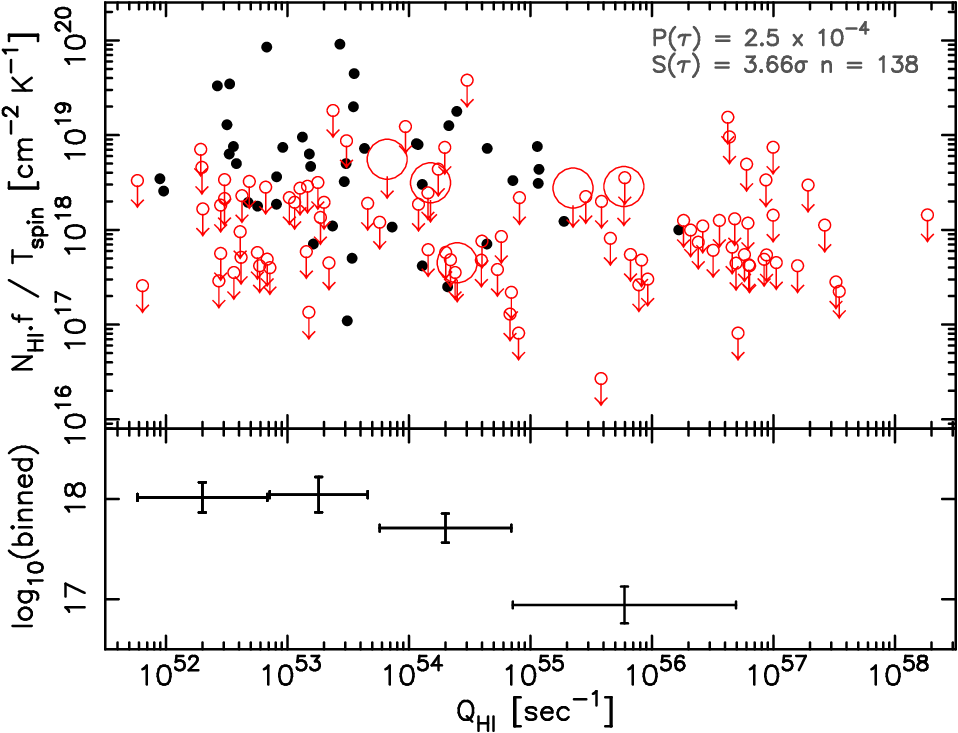}
\caption{The \HI\ 21-cm line strength versus the ionising photon rate of the $z\geq0.1$ associated absorption searches. As
  per usual (e.g. \citealt{cras16}), we include the \HI\ 21-cm non-detected $3\sigma$ limits as censored points via the {\em
    Astronomy SURVival Analysis} ({\sc asurv}) package \citep{ifn86}. A generalised non-parametric Kendall-tau test
  gives a probability of $P(\tau) = 2.53\times10^{-4}$ of the correlation occuring by chance, which is significant at $S(\tau)
  =3.66\sigma$, assuming Gaussian statistics. The inclusion of the forthcoming data (G17) increases the significance to
  $S(\tau) =5.74\sigma$ [$P(\tau) = 9.58\times10^{-9}$]. The bottom panel shows the binned values, including the limits
  via the Kaplan--Meier estimator, in equally sized bins. It is not possible to derive a value for the last bin since
  this comprises only limits. The horizontal error bars show the range of points in the bin and the vertical error bars
  the $1\sigma$ uncertainty in the mean value.}
\label{Q-N}
\end{figure} 
For example, the two detections at $z\gapp2.5$
(\citealt{mcm98,ubc91}, which have $Q_\text{\HI}\approx1\times10^{55}$~sec$^{-1}$, Fig.~\ref{N-ion}).
 Therefore, contrary to \citet{akk16}, we reaffirm that the luminosity--redshift degeneracy can be broken and that the
decrease in detection rate is caused by an increase in the UV luminosity.

Referring to the second panel of Fig.  \ref{all-rate}, from the sensitivities we expect a detection rate of $\gapp20$\%,
cf. 26\% for the whole sample, which rises to 29\% with the removal of the $Q_\text{\HI}\geq3\times10^{56}$~sec$^{-1}$
sources. This compares to the 25\% found by \citet{gss+06} and the 40\% by \citet{vpt+03}, although this was for compact
objects, which trace less UV luminous sources \citep{cw10,ace+12}.\footnote{The maximum expected detection rate is
  $\sim50$\% on account of the chance of the radiation being intercepted by the galactic disk along our sight-line (see
  \citealt{cw10} and references therein).} From a $\approx30$\% detection rate, we may expect $1.8\pm1.1$ (for
binomial statistics) detections out of our six targets. This is within $2\sigma$ of zero detections and so our results are not extraordinary.
Nevertheless, it would be of value to future surveys to analyse other possible factors which could affect the
detection of \HI\ 21-cm absorption.

\subsubsection{Dust reddening}

\begin{figure}
\centering 
\includegraphics[angle=270,scale=0.52]{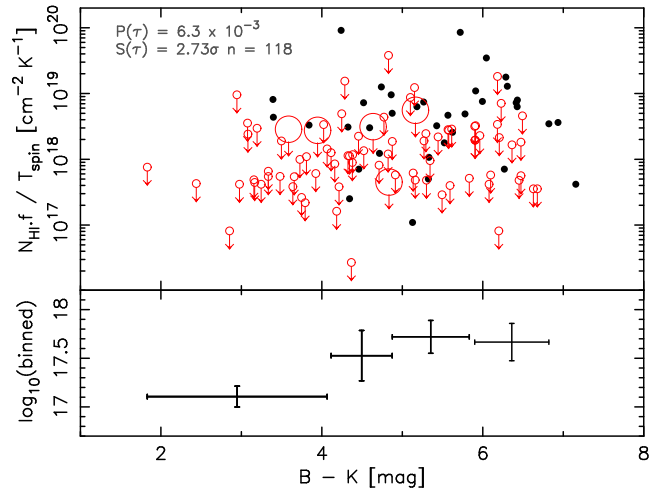}
\caption{The \HI\ 21-cm line strength versus the blue -- near-infrared colours of the $z\geq0.1$ associated absorption
  searches.  The significance rises to $3.80\sigma$ ($n=176$) with the inclusion of the forthcoming data (G17).}
\label{B-K}
\end{figure}

As discussed in Sect. \ref{intro}, the correlation between molecular fraction and optical -- near-infrared colour
\citep{cwc+11} is evidence of dust reddening, where the dusty sight-lines are more conducive to high molecular
abundance. From a sample of five red quasars searched in associated \HI\ 21-cm absorption, \citet{cmr+98} obtained four
detections, from which they suggested that optically selected samples, which have lower \HI\ 21-cm detection rates, may bias
against the detection of high column density absorbers. Following this, \citet{cwm+06} noted a trend between the \HI\ 21-cm
absorption strength and the optical -- near-infrared colour, with \citet{cw10} reporting a $3.63\sigma$ correlation
between $\int\tau dv$ and $V-K$ for the 58 searched associated absorbers for which the colours were available. Since we
have based the current sample on their faint blue-magnitudes, in Fig. \ref{B-K} we show the
current distribution in terms of $B-K$.
A generalised non-parametric Kendall-tau test gives a probability of $P(\tau) = 0.0063$ [$S(\tau) =2.73\sigma$] of the
correlation occuring by chance, with the inclusion of the forthcoming data (G17) increasing  the significance 
to $S(\tau) =3.80\sigma$ [$P(\tau) = 1.43\times10^{-4}$].  Furthermore, in the third panel of Fig.  \ref{all-rate} we see
a steep  increase in the detection rate with the red colour. This and the strong positive correlation with the line strength, indicates that
the reddening is caused by dust, the presence of which hinders excitation of the hydrogen above the lower hyper-fine ($F=0$)
ground state.

\subsubsection{Coverage of the background continuum flux}

From Equ. \ref{tau_obs}, we see that the observed optical depth can be reduced by low coverage ($f\ll1$) of the
background emission, an effect seen in both intervening and associated absorption (\citealt{cur12,cag+13},
respectively).  For a given (unknown) absorption cross-section, in the optically thin regime, the observed optical depth
will therefore be proportional to the extent of the radio emission (see \citealt{cmp+03}).  Extended emission may be
apparent through a steep spectrum, due to the radio jets being aligned normally to our
line-of-sight (seen edge-on), thus maximising the apparent size of the source. Another signature of extended emission
is a low/non-apparent turnover frequency, it being generally accepted that the turnover frequency of a radio source is
anti-correlated with its extent (e.g. \citealt{ffs+90}).

From a sample of near-by galaxies, \citet{cras16} found that, while there was a large overlap in the spectral indices of the
\HI\ 21-cm detections and non-detections, detections did tend to occur towards  sources
with higher turnover frequencies. 
\begin{figure}
\centering \includegraphics[angle=-90,scale=0.52]{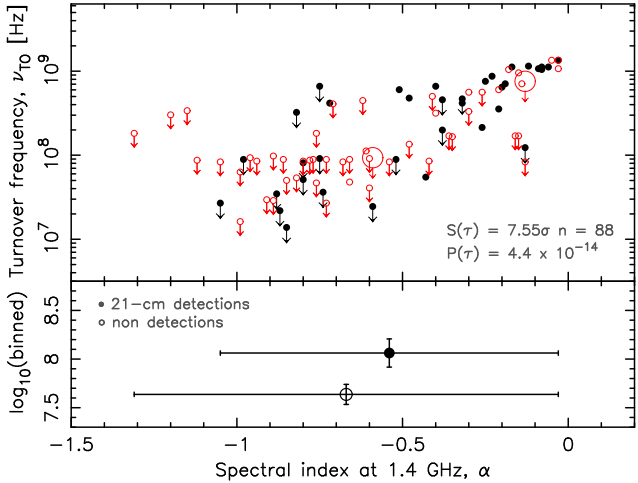}
\caption{The rest-frame turnover frequency versus the spectral index for the background sources for which these 
could be determined. If no turnover is apparent in the radio SED we assume that this occurs below the lowest
observed frequency (typically $\sim10$ MHz) and use this to assign an upper limit to $\nu_{_{\rm TO}}$.
Again, the filled symbols represent the detections and the unfilled the non-detections.
In the bottom panel, the binned values of the detections and non-detections are shown.}
\label{TO-SI}
\end{figure} 
We obtain the turnover frequencies and the spectral indices from the fits described in Sect.~\ref{ss} and 
plotting the \HI\ 21-cm absorption searches in Fig.~\ref{TO-SI}, we see a similar result where the detections have a mean turnover
frequency of $\left<\nu_{_{\rm TO}}\right>=115_{-33}^{+45}$~MHz, compared to $\left<\nu_{_{\rm TO}}\right>=43_{-9}^{+12}$~MHz, for the
non-detections. This is confirmed in Fig.~\ref{N-TO} (see also the fourth panel of Fig.~\ref{all-rate}), where the
positive correlation between the line strength and turnover frequency (suggesting an inverse correlation with source
size), indicates that the covering factor is important.
\begin{figure}
\centering \includegraphics[angle=-90,scale=0.6]{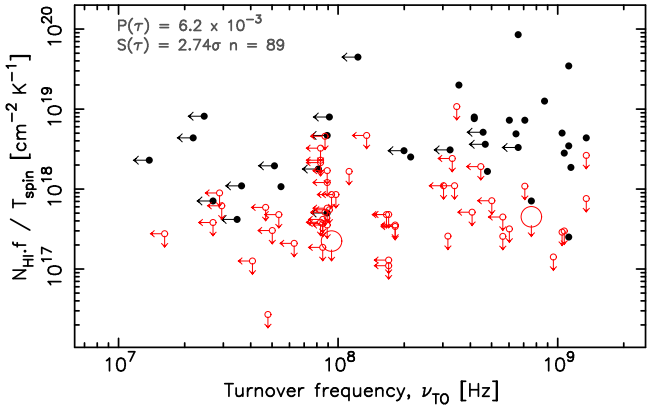}
\caption{The line strength versus the rest-frame turnover frequency. The inclusion of the forthcoming data (G17) does not change the significance ($n=100$, cf. 89 here).}
\label{N-TO}
\end{figure} 
However, as seen in the top panel of the figure, not having an evident turnover frequency does not
preclude a detection, with this being possible at $\nu_{_{\rm TO}}<20$~MHz. Therefore, it should be borne in mind that
the spectral properties offer only  an indirect measure of the emission region and the covering factor depends upon how
this is related to the absorption cross-section, as well as the alignment between absorber and emitter (see \citealt{cag+13}). 

\subsubsection{Summary}

In Table \ref{reasons} we summarise the probability of detection for the six targets discussed in this work, based upon
the above parameters.
These are obtained from the number of detections normalised by the total number of searches within the bin
which hosts the target (Fig. \ref{all-rate}). 
\begin{table} 
\centering
\begin{minipage}{80mm}
  \caption{The probabilities of detection for each of our six targets based upon the properties of the \HI\ 21-cm absorption searches.
    The rates are obtained from the intersection of the dotted vertical lines designating the values for our targets and the
    full line joining the binned values for all of the \HI\ 21-cm searches  (Fig. \ref{all-rate}).  
  The $B-K$ rates have been interpolated between $B-K = 3 \text{ and } 6.5$ and, as per the figure, these are intended as 
approximate guide values only.} 
\begin{tabular}{@{}l    c  c c  l c@{}} 
\hline
\smallskip
Source                & $z$-range &  $Q_\text{\HI}$&  Sensitivity & $B-K$  & $\nu_{\rm TO}$\\
\hline
0114+074       & 0.336--0.356      & ---   & 27\%     & ---  & 33\%\\  
0137+012      &  0.252--0.269        &  31\%   & 22\%     &  22\%  & ---\\  
0240--217     & 0.306--0.324        &   41\%   & 26\%     & 27\%  &   52 \%\\           
  0454+066     & 0.396--0.417              &  28\%   & 24\%   &   20\%  & ---\\             
1456+044    &  0.382--0.400         &  44\%    & 24\%     &  26\%  & ---\\                                            
1509+022    &  0.216--0.232           &   52\%  & 18\%     &   29\% & ---\\                       
\hline       
\end{tabular}
\label{reasons}  
\end{minipage}
\end{table} 
From these we see that, on an individual basis, the odds are against a detection with a maximum probability of 52\% for a
single parameter ($Q_\text{\HI}$ for 1509+022 and $\nu_{\rm TO}$ for 0240--217). This, however, will be tempered by the probability from the other parameters, which 
are not independent\footnote{For example, the relationship between $Q_\text{\HI}$ (the UV luminosity) and the
  $B-K$ colour, where $B$ will be anti-correlated with $L_{\rm UV}$, particularly at $z\gapp3$. Since the radio and UV
  luminosities are correlated \citep{cw10}, there will also be a relationship between $Q_\text{\HI}$ and the search sensitivity.},
making the overall probability difficult to ascertain for each source.

So in addition to our hypothesis that we only expect detections where $L_{\rm UV}\lapp10^{23}$ \WpHz\
($Q_\text{\HI}\lapp3\times10^{56}$~sec$^{-1}$), detection rates may be maximised through the selection of red,
gigahertz peaked spectrum sources.  The colour is indicative of a dusty environment, more conducive to the presence of
cold neutral gas, and the turnover frequency indicative of an increased covering factor, due to the compact radio
emission. If the projected extent of the radio emission is dictated by the orientation of the jets along the sight-line,
this presents an issue for the model where the absorption occurs in the obscuring torus invoked by unified schemes of
active galactic nuclei (e.g. \citealt{jm94,cb95,mot+01,pcv03,gss+06,gs06a}), since this is less likely to intercept the
sight-line in compact objects (e.g. \citealt{gss+06}). However, as well as occuring in outflowing gas
(e.g. \citealt{vpt+03,moe+03,mot+05,mto05,mhs+07,asm+15a}), the bulk of the absorption is believed to occur in the
large-scale galactic disk, rather than the sub-pc torus\footnote{Although this does make a small contribution
  (\citealt{cdda16}).}, which is expected to have a random orientation to the radio jets (\citealt{cw10} and references therein).

\section{Conclusions}
\label{sec:conc}

We have undertaken a survey for associated \HI\ 21-cm and OH 18-cm absorption in six $z\approx0.2-0.4$ radio sources
with the GMRT.  Despite selecting targets which have ionising photon rates lower than the critical value, above which
the neutral gas is believed to be completely ionised, and reaching the sensitivities of previous detections, we do not
detect either transition. Given that the OH absorption strength is expected to be $\lapp10^{-4}$ times that of the
\HI\ 21-cm absorption \citep{cdbw07}, the non-detection of this radical is  not surprising, 
especially in the knowledge that our targets are not as reddened as originally believed. Regarding the
\HI, by combining our results with previous searches, in addition to the ionising photon rate (UV luminosity), we find correlations between
 the detection rate and:
\begin{enumerate}
\item The blue -- near-infrared colour, which is indicative of the
reddening being due to dust, the presence of which helps in maintaining a low gas spin temperature in the case of the detections. 
\item The turnover frequency of the radio SED of the source, which is believed to be anti-correlated with the size of the radio 
emission (e.g. \citealt{ffs+90}). This itself is anti-correlated with the line strength, due to lower coverage of the measured flux,
and so lower optical depths are expected for extended sources \citep{cag+13}.
\end{enumerate}
The detection rate is, of course, also subject to the search sensitivity with a maximum rate of $\approx30$\% for
$N_{\rm HI}\lapp10^{18}\,(T_{\rm spin}/f)$ \scm. A maximum of $\sim50$\% may be expected, due to the
orientation of the absorbing gas with respect to the continuum emission \citep{cw10}, and so 
in addition to maximising the
sample size, future surveys should also target optically faint/obscured, radio bright
objects. Since this could preclude objects with measured optical redshifts, full band ``spectral scans'' towards sufficiently
bright radio sources may be the way forward (e.g. \citealt{cwmk03,asm+15}).

We confirm that associated \HI\ 21-cm absorption has never been detected above the
theoretical value of $Q_\text{\HI}\approx3\times10^{56}$~sec$^{-1}$ \citep{cw12}. From the highest reliable detected value of
$Q_\text{\HI}=1.7\times10^{56}$~sec$^{-1}$ (a monochromatic $\lambda = 912$~\AA\ luminosity of $L_{\rm
  UV}\approx2\times10^{23}$ \WpHz), the binomial probability of the observed 0 detections out of 29 searched above this
value is just $4.80\times10^{-7}$, which is significant at $5.03\sigma$.  Adding, the forthcoming results of
\citet{gdb+15}, the probability of the 0 detections out of 63 searches is just $2.58\times10^{-11}$, which is significant
at $6.67\sigma$. Hence, the case for a critical ultra-violet luminosity is
strengthened. Given the unbiased (i.e. inhomogeneous) nature of the whole sample, this $L_{\rm UV}\sim10^{23}$ \WpHz\
($Q_\text{\HI}\approx3\times10^{56}$~sec$^{-1}$) appears to be universal and the selection of sources of known optical
redshift leads to the selection of higher luminosities at high redshift (see \citealt{msc+15}), where all of the gas is
ionised. That is, even the Square Kilometre Array will not detect \HI\ 21-cm absorption in these. This leads us to reiterate
that spectral scans towards optically faint objects is the best strategy and this is where the SKA will excel due to its
wide instantaneous bandwidth.

\section*{Acknowledgements}

We wish to thank the anonymous referee for their helpful comments, as well as Katie Grasha and Jeremy Darling for a
draft manuscript of their forthcoming paper.  This research has made use of the NASA/IPAC Extragalactic Database (NED)
which is operated by the Jet Propulsion Laboratory, California Institute of Technology, under contract with the National
Aeronautics and Space Administration.  This publication makes use of data products from the Wide-field Infrared Survey
Explorer, which is a joint project of the University of California, Los Angeles, and the Jet Propulsion
Laboratory/California Institute of Technology, funded by the National Aeronautics and Space Administration.  This
publication makes use of data products from the Two Micron All Sky Survey, which is a joint project of the University of
Massachusetts and the Infrared Processing and Analysis Center/California Institute of Technology, funded by the National
Aeronautics and Space Administration and the National Science Foundation. GALEX is operated for NASA by the California
Institute of Technology under NASA contract NAS5-98034.  This research has also made use of NASA's Astrophysics Data
System Bibliographic Services and {\sc asurv} Rev 1.2 \citep{lif92a}, which implements the methods presented in
\citet{ifn86}.


\label{lastpage}

\end{document}